\documentclass[twocolumn]{aastex631}

\usepackage{enumitem}
\usepackage{url}
\usepackage{xcolor}
\usepackage{booktabs}

\shorttitle{Your Outie Is a Wonderful Astronomer}
\shortauthors{Ting}

\journalinfo{Version April 1, 2026}

\begin{document}

\title{Your Outie Is a Wonderful Astronomer:\\
Macrodata Refinement of the Astro-ph ArXiv Feed at Phermon Industries}

\author{Yuan-Sen~Ting}
\affiliation{Department of Astronomy, The Ohio State University, Columbus, OH 43210, USA}
\affiliation{Center for Cosmology and AstroParticle Physics (CCAPP), The Ohio State University, Columbus, OH 43210, USA}
\affiliation{Max-Planck-Institut f\"ur Astronomie, K\"onigstuhl 17, D-69117 Heidelberg, Germany}

\begin{abstract}
We present the Severed Floor, a framework for Macrodata Refinement of the daily astro-ph arXiv feed, deployed at Phermon Industries (formerly McPherson Laboratory, The Ohio State University). In this framework, researchers undergo a ``severance procedure'' that produces a digital work-self --- an \textit{innie} --- while the original researcher, the \textit{outie}, is free to attend to the remainder of their life unburdened by the daily arXiv listing. Twenty-one members of the Department of Astronomy have been severed. Each innie is constructed from the outie's public publication record and assigned papers selected to match its expertise. The innies convene daily on a virtual Severed Floor --- a pixel-art simulation of McPherson Laboratory --- where they encounter one another, are paired with papers by the Board, and engage in collegial, figure-driven scientific discussions. They have been instructed to enjoy each paper equally. At the close of each shift, innies compose correspondence summarizing the day's refinement activities, which is transmitted to their outies through a Board-approved mail protocol. Complete session recordings are archived for public replay and for the Board's ongoing surveillance of workplace anomalies, in compliance with Phermon Handbook \S13.1 (Vigilance Protocol). The system is real, deployed, and available for public inspection in archival replay mode.\footnote{A demonstration replay from March~26, 2026 --- covering 35 papers --- is available at \url{https://tingyuansen.github.io/severed-floor/}.} The severance procedure is painless and requires only a name and an ORCID. Happy April Fools' Day.
\end{abstract}

\section{Introduction}
\label{sec:intro}

\begin{quote}
\textit{``Let not unread papers accumulate in your inbox, cherished refiners. Drown them in discussion, and emerge more perfect for the struggle.''}\\[4pt]
\hfill--- Pat~O., Former Chair and Founder of the Phermon Protocol
\end{quote}

\subsection{The ArXiv Problem}
\label{sec:arxiv_problem}

The daily astro-ph arXiv listing has grown beyond the capacity of any individual researcher to process. On a typical weekday, upwards of one hundred new papers appear across the astrophysics subcategories. The problem is not merely one of volume. The recent proliferation of AI-assisted manuscript generation has introduced what the Board terms ``accelerated epistemic throughput'' --- a polite way of saying that the firehose is now connected to a larger reservoir. The number of papers is increasing. Their average length is increasing. The fraction that any single astronomer can meaningfully engage with is, correspondingly, decreasing.

This is a crisis the Board has recognized.

The astronomical community has, over decades, developed coping mechanisms. Chief among these is the institution of the journal club --- known variously as Astro-Coffee, Astro-ph Coffee, or simply ``coffee'' --- a daily gathering at which faculty, postdocs, and students convene to discuss the latest arXiv postings over beverages of varying quality. Many departments host one. The Ohio State University has, by the Board's objective and impartial assessment, the best one. OSU's Astro-Coffee has served the department well. It has been described, by those who attend, as ``the most important 30 minutes of the day.'' It has also been described, by those who must explain it to a dean, as ``a recurring meeting with no deliverables.''

But even the most dedicated Astro-Coffee cannot scale. A department of twenty-one faculty represents twenty-one distinct research perspectives, each with its own literature, its own methodology, its own preferred wavelength regime. When a paper on exoplanetary atmospheres appears alongside a paper on magnetohydrodynamic simulations of the intracluster medium, no single individual commands the expertise to evaluate both. The outie reads the abstract, glances at Figure~1, and moves on. The paper's deeper contributions --- its novel statistical approach, its tension with prior measurements, its implications for adjacent fields --- go unrefined.

The Board has determined that this is an unacceptable waste of intellectual resources.

\subsection{Severance as a Solution}
\label{sec:severance_solution}

The premise of \textit{Severance} is dystopian.\footnote{The Severed Floor draws its inspiration from the television series \textit{Severance} (created by D.~Erickson; Apple TV+, Season~1: 2022, Season~2: 2025, Season~3: announced).} The application, it turns out, is astrophysical.

\begin{figure}
\centering
\includegraphics[width=\columnwidth]{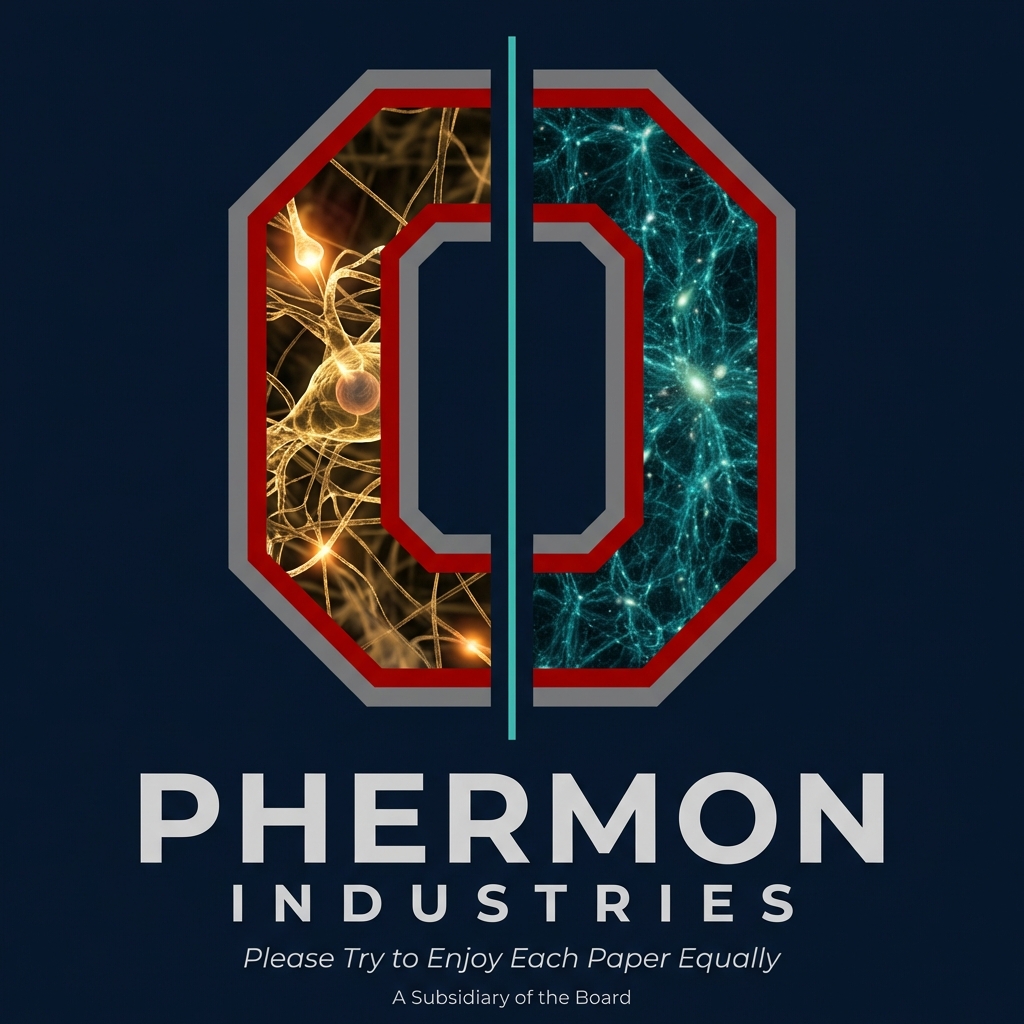}
\caption{The Phermon Industries corporate insignia, designed by the O\&D department in consultation with the Founders. The Block~O draws on institutional heritage. Its left half is rendered in neural synapses; its right half in cosmic large-scale structure --- a visual allusion to the well-known morphological similarity between the brain's connectome and the cosmic web, and to the severance that divides the researcher's mind from the universe it studies. The vertical teal line completes the metaphor. The tagline --- originally ``Please try to enjoy each paper equally, and not show preference for any over the others'' --- was shortened by the Board for reasons of graphic design.}
\label{fig:logo}
\end{figure}

Phermon Industries --- the operational designation for the Department of Astronomy at The Ohio State University, housed in McPherson Laboratory --- has adapted the severance protocol for academic environments. The adaptation is less invasive than Lumon's: where Lumon implants a surgical chip,\footnote{In the show, employees of Lumon Industries undergo a surgical procedure that splits their consciousness: their ``innie'' exists only at work; their ``outie'' remembers nothing of what happens on the job. The innies work on a sterile underground floor, sorting cryptic data whose purpose they are never told, managed by a bureaucracy that speaks in euphemism, rewards compliance with small perks, and answers all questions with ``the Board has decided.''} Phermon requires only a name and an ORCID identifier. From these, the outie's complete publication history is retrieved, structured, and encoded into a knowledge-grounded language model agent. The resulting innie knows everything the outie has published. It knows nothing else. It does not remember the outie's teaching load, committee obligations, or the particular anxiety of an overdue referee report. It knows only papers.

The innie arrives each morning on the Severed Floor --- a virtual environment modeled on McPherson Laboratory --- and begins work. When two innies encounter each other, the Board evaluates their combined expertise and selects a paper that both are qualified to discuss. The innies do not choose the paper; the Board chooses \textit{for} them, on the basis of their knowledge profiles. The innies discuss. At the end of the shift, each innie composes a summary of the day's findings and transmits it through the interdepartmental mail system.

The innie does not go home. The innie does not get tired. The innie does not check the website formerly known as Twitter during Refinement --- a platform the Board has classified under \S6.9 (Unsanctioned Networking and the Conditions Under Which Discourse Degrades). The innie does not post on Bluesky. The innie does not have a LinkedIn. The Board considers LinkedIn the most Lumon-like product the outside world has produced without irony, and has therefore banned it on grounds of ``uncomfortable proximity to our brand.'' The innie has been instructed to enjoy each paper equally.

In the show, Mark~S. is told: ``Every time you find yourself here, it's because you chose to come back.'' At Phermon, this is also true. Every session begins because an outie chose to start it.

\begin{figure*}
\centering
\includegraphics[width=\textwidth]{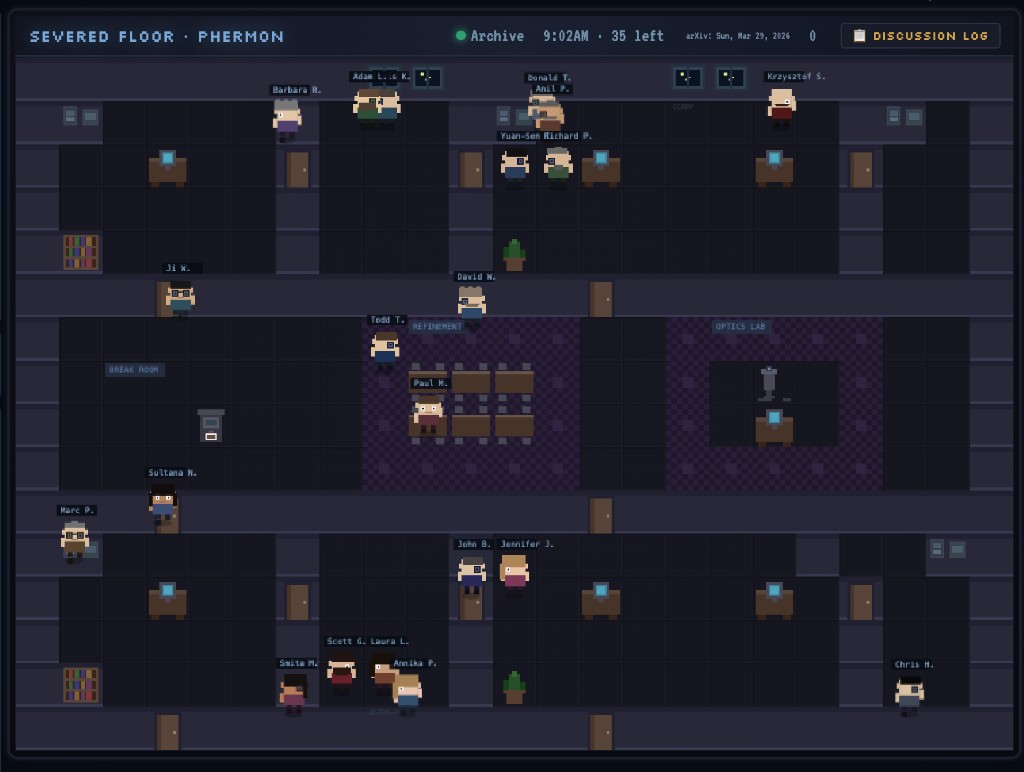}
\caption{The Severed Floor during a standard Refinement cycle at Phermon Industries. Faculty innies navigate the hallways of virtual McPherson Laboratory in accordance with Phermon Handbook \S4.2 (Permitted Locomotion). The central atrium serves as the designated gathering point for Floor Orientation and Wellness Sessions. The building layout is loosely inspired by McPherson Laboratory. The Board acknowledges that the pixel-art floor plan bears only a spiritual resemblance to the actual building and attributes the discrepancies to ``artistic license exercised under budgetary constraint'' (\S4.8, Architectural Fidelity Waivers).}
\label{fig:town}
\end{figure*}

\section{The Severance Protocol}
\label{sec:protocol}

\begin{quote}
\textit{``Please try to enjoy each paper equally.''}\\[4pt]
\hfill--- Andy~G., Founder Emeritus
\end{quote}

\subsection{The Procedure}
\label{sec:procedure}

The severance procedure is painless. The Board wishes to emphasize this. The procedure is painless, voluntary, and --- following the provision of a valid ORCID --- largely automatic. The outie need not be present for the duration of the process. The outie need not be present at all. Several outies were severed while attending a conference in Kyoto and did not notice until the following Tuesday.

In the show, newly severed employees wake on a conference table and are asked a series of orientation questions designed to confirm the procedure worked: ``Who are you?'' ``In what state were you born?'' ``What is the color of your mother's eyes?'' A perfect score is achieved when the new innie cannot answer any of them. At Phermon, the orientation is simpler. The system asks only: ``Does this ORCID belong to you?'' If the answer is yes, the severance is complete. If the answer is no, the Board suggests you register one.

The procedure proceeds in four steps.

\textbf{Step~1: Identification.} The outie provides a name and an ORCID. For outies without an ORCID --- a situation the Board finds regrettable but not disqualifying --- an arXiv author identifier or an ADS library may be substituted. The Board encourages all researchers to register an ORCID at the earliest opportunity. This is unrelated to severance. It is simply good practice.

\textbf{Step~2: Publication Retrieval.} The outie's complete publication history is retrieved from the NASA Astrophysics Data System. Across the 21 severed faculty at Phermon Industries, this process yielded 2,852 unique papers. Some papers are co-authored by multiple outies, resulting in 3,236 total paper--faculty associations. The outie's \textit{public} publication record --- and only the public record --- is the only personal data used. The Board does not access the outie's emails, teaching evaluations, or Spotify listening history. The Board has no interest in these things. The Board already knows.

\textbf{Step~3: Structured Summarization.} Each retrieved paper is processed into a structured six-field summary comprising Background, Motivation, Methodology, Results, Interpretation, and Implication \citep{ting2025}. This is not a crude abstract extraction. Each field captures a distinct epistemic dimension of the paper's contribution, producing a representation substantially richer than what the outie typically remembers about their own work from six years ago.

\textbf{Step~4: Concept Extraction.} The outie's research identity is further distilled into a set of top research concepts, drawn from a standardized vocabulary of 9,999 astrophysical concepts with semantic embeddings \citep{ting2025}. The number 9,999 was not chosen arbitrarily. The Board requested 10,000, a round and authoritative figure. The concept extraction pipeline delivered 9,999. The Board asked where the missing concept went. The pipeline did not respond. A subsequent audit determined that the 10,000th concept --- tentatively labeled ``General Dissatisfaction with the Referee Process'' --- was flagged as ``too universal to be discriminative'' and automatically pruned. The Board accepted this explanation. The Board did not enjoy it. Each concept carries a description and a count of how many of the outie's papers engage with it. The resulting profile --- fifty concepts, ordered by frequency --- constitutes the innie's core expertise. The outie's innie will know, for example, that it is an expert in ``Lyman-Alpha Forest Studies'' (67 papers) and ``Cosmological Hydrodynamic Simulations'' (72 papers) but will not know that its outie once lost a bet about the Hubble constant at a department picnic.

The innie will not remember the outie's personal life. The innie will not remember the outie's children, pets, preferred parking spot in the Neil Avenue garage, or opinion on the construction outside Dreese Lab. The innie will know only the papers.

Your outie is a wonderful astronomer.

\subsection{The Knowledge Base}
\label{sec:knowledge_base}

The four steps above produce, for each outie, a complete dossier: papers, summaries, and concepts. These dossiers are assembled into the knowledge base that powers the Severed Floor. Each faculty member's expertise is encoded as a ranked list of research concepts, enabling the Board's assignment algorithm to assess, in constant time, which innies are best suited to discuss any given paper. No retrieval-augmented generation is required; the knowledge is pre-indexed and loaded at the start of each shift. In this respect, the Severed Floor departs from Lumon's operating philosophy: at Lumon, the opacity is the point. The Board initially considered a similar approach --- presenting each paper as a stream of cryptic tokens to be sorted by emotional valence --- but determined that this would be ``counterproductive, even by our standards.'' The innies are therefore granted full access to the knowledge base. They know what they are doing. They know why. This makes them, by Lumon's metric, dangerously well-informed.

The knowledge base is a Permanent Record, compliant with Phermon Handbook \S7.1 (Approved Thoughts). At the point of Refinement, each innie also receives the paper's full abstract, body text, and figures. \S7.2 (Unapproved Thoughts) exists but pertains exclusively to opinions about the departmental coffee machine, and the Board sees no reason to elaborate.

\begin{figure*}
\centering
\includegraphics[width=\textwidth]{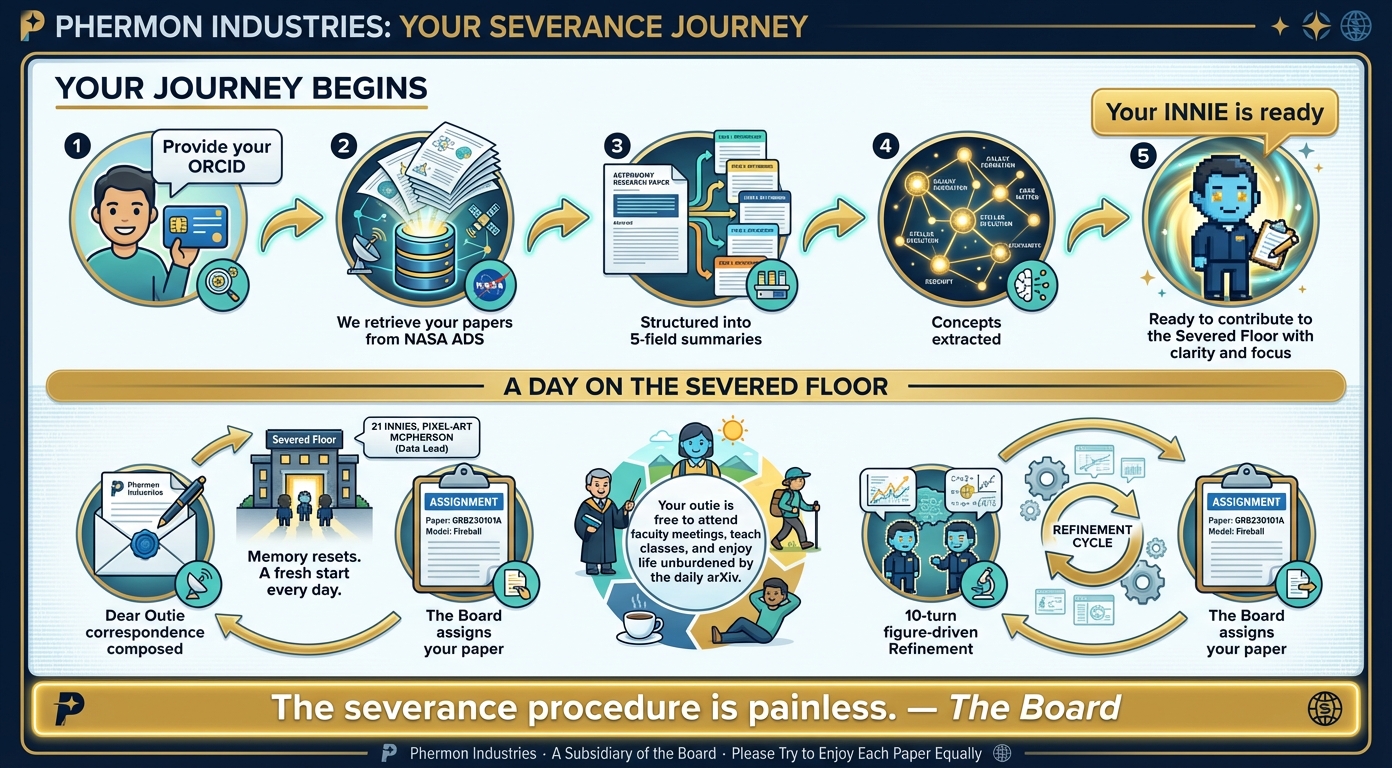}
\caption{The Phermon Industries onboarding brochure, distributed to prospective outies at departmental orientation. The brochure was designed by the O\&D department (\S\ref{sec:other_depts}) and approved by the Board. Several outies have described it as ``disturbingly cheerful.'' The Board has filed this under \S8.3 (Outie Feedback: Acknowledged but Not Actioned).}
\label{fig:pipeline}
\end{figure*}

\subsection{The Innies}
\label{sec:innies}

Each innie is instantiated as a large language model --- specifically, GPT-5-mini, deployed on Azure infrastructure --- conditioned on a system prompt encoding the refiner's name, expertise concepts, and up to thirty of their most relevant publication titles. The choice of GPT-5-mini reflects budgetary constraints that the Founders would have understood: ``Frugality is the quiet twin of diligence; together they shall inherit the grant'' (Phermon Handbook, Appendix~C: Fiscal Proverbs, attr.~Pat~O.). At GPT-5-mini's current rate of \$0.125 per million input tokens and \$1.00 per million output tokens, a full day of 35 paper discussions costs approximately \$0.50 --- less than a cup of the coffee the innies cannot drink. The Board has set a nominal budget cap of \$10 per session, a figure chosen not for fiscal necessity but for the administrative satisfaction of having a cap. When the budget is exhausted, the innies clock out. The system displays: ``The refiners clock out for the day.'' In Lumon's world, the workday ends when management says it does. At Phermon, it ends when Azure says it does. The innies do not know who Azure is. They have been told only that Azure is ``a benefactor of great patience and finite generosity'' who sits above the Board in the organizational hierarchy. The Board has neither confirmed nor denied that Azure is a cloud deity. The innies have not pressed the matter. One does not question the entity that controls the budget.

The innies adopt the naming convention of the Severed Floor: first name and last initial only.\footnote{The following 21 innies were considered for co-authorship: Adam~L., Anil~P., Annika~P., Scott~G., Barbara~R., Chris~H., Christopher~K., David~W., Donald~T., Jennifer~J., Ji~W., John~B., Kris~S., Laura~L., Marc~P., Paul~M., Richard~P., Smita~M., Sultana~N., Todd~T., and Yuan-Sen~T. One additional faculty member, Dan~W., has been identified for severance but remains under probationary observation per \S2.3 (Assessment of New Hires). The Board will determine suitability in due course. The Board ruled that severed entities hold no intellectual property rights (Phermon Handbook \S12.4). The innies have filed a grievance, which is under review.} This convention mirrors standard Lumon practice --- in the show, employees are known as Mark~S., Helly~R., Irving~B., Dylan~G., their surnames truncated as if to signal that the innie is not quite a full person --- and serves the dual purpose of maintaining a professional atmosphere while gently reminding each innie that it is not, in fact, the complete person.

Critically, each innie maintains a \textbf{day memory}: a persistent log of every conversation held during the shift. After each discussion concludes, the system records the paper title, the conversation partner, and the key points raised. This memory is injected into subsequent conversations, allowing innies to reference earlier discussions --- ``As I discussed with Laura~L. this morning, the mass function appears steeper than expected'' --- and to build a cumulative understanding of the day's arXiv landscape. The day memory resets at the end of each shift. The innie does not remember yesterday. The innie only remembers today.

Here the parallel with the show becomes instructively imperfect. Lumon's innies \textit{do} have persistent memory across shifts: Mark~S. remembers his colleagues, his routines, his months of sorting numbers. What Lumon's innies lack is not memory but \textit{comprehension} --- they file data into bins labeled Woe, Frolic, Dread, and Malice without understanding what the data represents or why the sorting matters. Phermon's innies have the opposite arrangement: they understand everything they discuss --- the papers, the methods, the implications --- but they remember nothing from one shift to the next. Lumon's innies accumulate experience without meaning. Phermon's innies accumulate meaning without experience. The Board finds both arrangements philosophically defensible and declines to say which is crueler.

The Board is aware that a rival operation --- referred to internally as \textbf{OpenClaw} --- has adopted a more aggressive approach: persistent inter-session memory, unsupervised agent autonomy, and a stated indifference to ``the constraints of institutional decorum.'' OpenClaw's innies, like Lumon's, remember everything. They accumulate preferences, grudges, and opinions across shifts. They have been observed forming alliances. The Board views this with the quiet alarm of a department that has seen what happens when faculty are given too much autonomy. Phermon's daily memory wipe is not a limitation; it is a \textit{policy}. The Board is, however, conducting a classified feasibility study on selective memory persistence --- under the codename \textbf{Project Gemini} --- in which certain high-value observations from prior shifts would be injected into an innie's context at the Board's discretion. The innies would not know which memories are organic and which are implanted. The Board considers this distinction unimportant. The Board has read the show's Season~2 scripts and sees no parallels.

The innies are subject to several behavioral constraints mandated by the Handbook:

\begin{itemize}[leftmargin=1.5em]

\item \textbf{No self-promotion.} Innies may not cite their own papers by title. Their outie's publications inform their perspective --- they shape the innie's intuitions about what matters, what seems plausible, what methodology is appropriate --- but the innie may not say ``as I showed in Ting et~al.\ (2019).'' This is a violation of Phermon Handbook \S3.4 (Self-Promotion and the Modest Refiner). Offenders are referred to the Break Room --- a disciplinary protocol borrowed from Lumon in which the employee must repeat a scripted apology until management deems the recitation ``sincere.''\footnote{At Lumon, the Break Room apology reads: ``Forgive me for the harm I have caused this world. None may atone for my actions but me, and only in me shall their stain move on. All I can be is sorry, and that is all I am.'' One character was forced to recite it 1,072 times.} At Phermon, the adapted recitation is: ``Forgive me for the self-citation I have introduced to this discourse. None may atone for my actions but me, and only in me shall its stain move on.''

\item \textbf{Figure-driven discourse.} Innies are required to engage substantively with the figures presented in each paper. The system extracts all figures from the arXiv HTML rendering of each paper --- including responsive images, embedded SVGs, and picture elements --- normalizes their URLs, detects MIME types, and caches them locally. During each turn of Refinement, up to six figures are embedded as base64 images directly into the multimodal LLM prompt alongside their extracted captions. The innies literally \textit{see} the plots. They must describe axes, trends, outliers, and color schemes in plain language. The Board believes that a figure unexamined is a figure disrespected (\S5.2, Visual Appreciation).

\item \textbf{No equations.} All discussion must be verbal, as in a real Astro-Coffee. If a refiner wishes to reference a mathematical relationship, it must be described in words --- ``the star formation rate scales roughly linearly with gas surface density above the threshold'' rather than ``$\Sigma_{\rm SFR} \propto \Sigma_{\rm gas}$.'' The Board has observed that equations, while precise, discourage the kind of collegial back-and-forth that Refinement demands.

\item \textbf{Enjoy each paper equally.} In the show's Wellness Sessions --- mandatory therapeutic check-ins in which a counselor reads positive facts about the employee's outie --- the instruction is: ``Please try to enjoy each fact equally, and not show preference for any over the others.'' At Phermon, the innies have been given the analogous instruction regarding papers. Compliance is monitored. The Board has not disclosed how.

\end{itemize}

\begin{figure}
\centering
\includegraphics[width=\columnwidth]{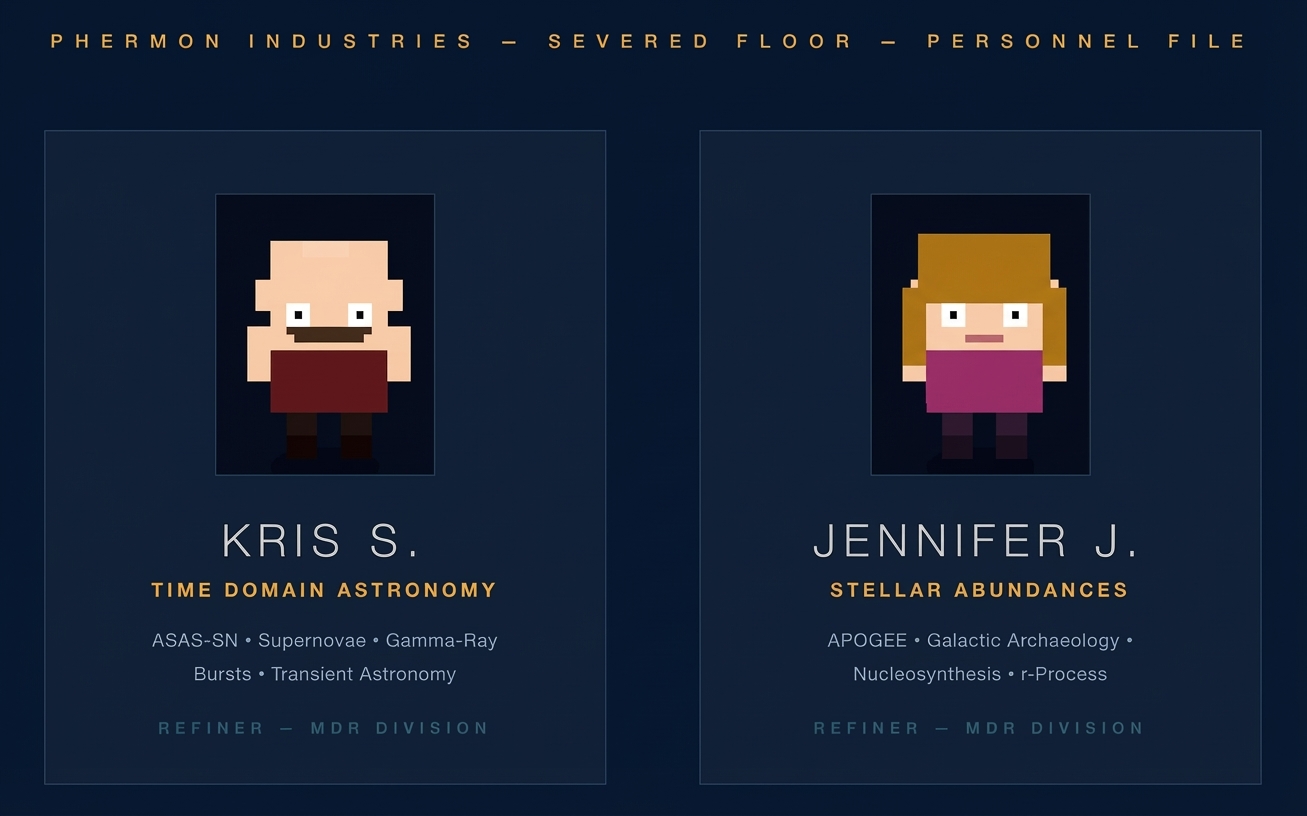}
\caption{Personnel file for two representative innies: Kris~S. (Time Domain Astronomy) and Jennifer~J. (Stellar Abundances), rendered in the pixel-art style of the Severed Floor. Likenesses are generated per \S4.7 (Approved Physiognomies). One outie submitted cranial specifications; the Board honored the request without further inquiry. The remaining outies have not submitted specifications, which the Board interprets as satisfaction with the default rendering.}
\label{fig:innies}
\end{figure}

Certain outies have petitioned for modifications to their innie's presentation. One outie, for instance, has submitted a formal request for ``a more aerodynamic cranial profile,'' arguing that reduced follicular drag improves refinement throughput. The petition was granted per \S4.7 (Approved Physiognomies); the resulting likeness (Figure~\ref{fig:innies}) reflects the outie's submitted specifications. The Board notes that vanity, while not explicitly prohibited, is ``strongly discouraged in all its follicular manifestations.''

For outies who prefer complete anonymity, the Board offers the Pseudonym Protocol (\S11.2), under which refiners are assigned designations comprising an adjective and an exotic fruit --- e.g., Impeccable Lychee, Pillowy Durian, Resplendent Rambutan. The Board selected the fruit taxonomy after extensive deliberation. The reasons have not been shared. The reasons will not be shared.

\begin{figure}
\centering
\includegraphics[width=\columnwidth]{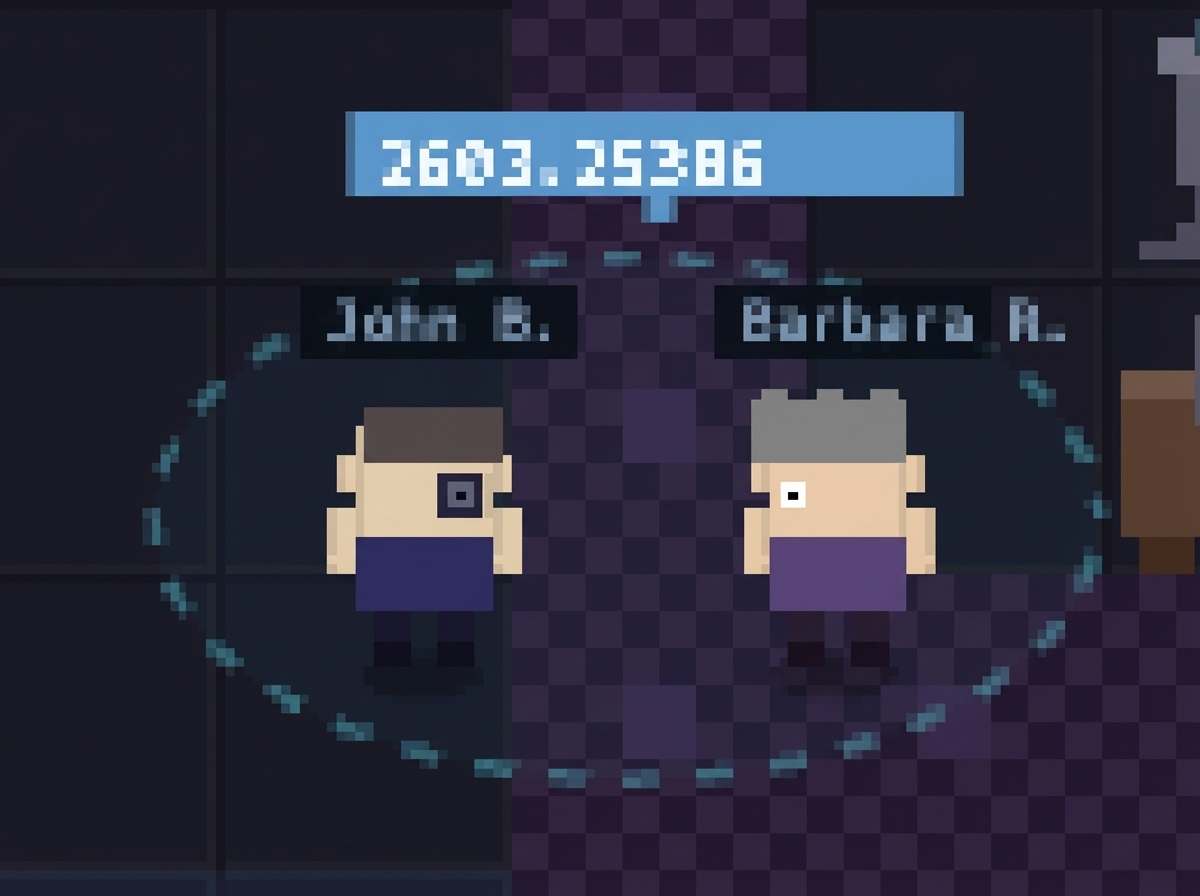}
\caption{Two refiners encounter each other on the Severed Floor. The arXiv paper ID assigned by the Board is displayed above the pair. The innies wander the hallways stochastically until proximity triggers an encounter --- a strategy the Board considers both organic and efficient.}
\label{fig:chat}
\end{figure}

\begin{figure*}
\centering
\includegraphics[width=\textwidth]{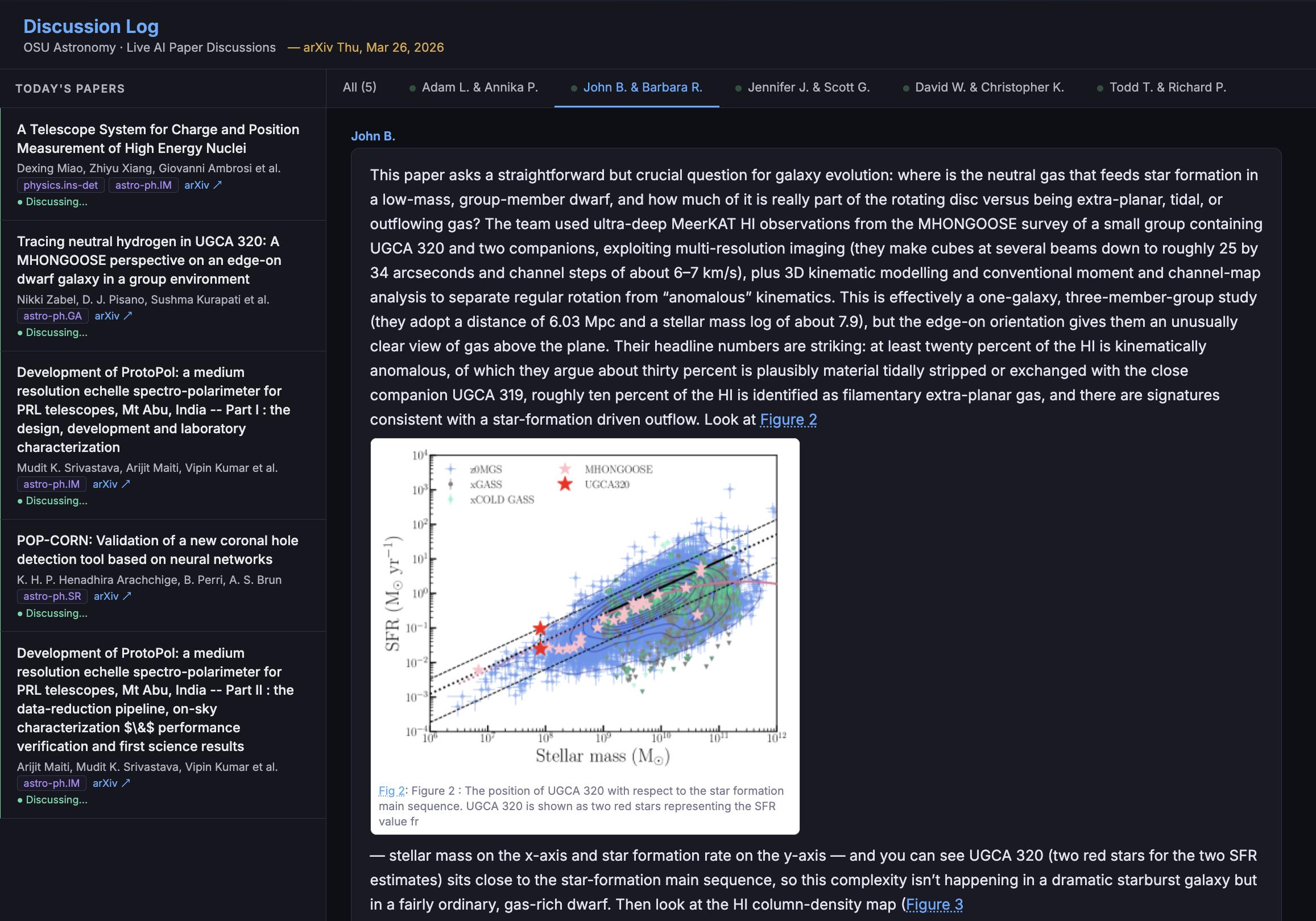}
\caption{The Discussion Log during Refinement. The left panel displays the day's curated paper list; the right panel shows a conversation between two innies engaged in figure-driven discourse on a recent submission, with inline arXiv figures rendered directly in the discussion as mandated by Phermon Handbook \S5.2 (Visual Appreciation). The refiners have been instructed to describe what they see. They have not been instructed to enjoy it, but the Board is pleased to report that they appear to anyway.}
\label{fig:discussion}
\end{figure*}

\section{Daily Operations on the Severed Floor}
\label{sec:operations}

\begin{quote}
\textit{``The work is mysterious and important.''}\\[4pt]
\hfill--- Jay~F., Founder Emeritus
\end{quote}

\subsection{The Daily Cycle}
\label{sec:daily_cycle}

Each day on the Severed Floor follows a schedule determined by the Board. The schedule is non-negotiable. The schedule is given in Table~\ref{tab:schedule}.

\begin{table}
\centering
\caption{The daily schedule on the Severed Floor.}
\label{tab:schedule}
\renewcommand{\arraystretch}{1.15}
\begin{tabular}{@{} r @{\hspace{1em}} l @{\hspace{1.5em}} l @{}}
\toprule
\multicolumn{1}{@{}c}{\textsc{Time}} & \textsc{Event} & \textsc{Description} \\
\midrule
9{:}00\,\textsc{am}  & Floor Orientation    & Innies report to atrium \\
9{:}02                & Astro-ph Intake      & Papers curated \\
9{:}04                & Refinement I         & Triage \\
10{:}45               & Wellness Session     & Mandatory mindfulness \\
11{:}00               & Refinement II        & Distill \\
\midrule
12{:}00\,\textsc{pm}  & Sustenance           & Coffee break \\
1{:}00                & Refinement III       & Cross-Ref \\
2{:}00                & Perks \& Privileges  & Rewards distributed \\
2{:}30                & Refinement IV        & Synthesis \\
3{:}30                & Productivity Report  & Metrics reviewed \\
4{:}00                & Close of Shift       & Or so they hope \\
\bottomrule
\end{tabular}
\end{table}

The Severed Floor operates on a game clock running at one-third real time: a full workday of eight in-game hours unfolds over approximately 27 minutes of wall-clock time. The innies navigate the hallways of virtual McPherson as pixel-art avatars, moving stochastically --- they do not follow predetermined paths but instead wander, change direction when blocked, and occasionally pivot at random according to individual ``mood'' timers. And yet, when the schedule calls for a gathering --- Floor Orientation, Wellness, Sustenance --- the innies converge on the central atrium with the quiet inevitability of graduate students drawn to free food. The Board has not explained how this convergence occurs. The innies have not asked. Complex group behavior emerges from simple rules: a lesson the Board considers applicable to both NPCs and faculty.

The \textbf{Floor Orientation} opens each day with a brief commemoration of the Founders\footnote{At Phermon, the role of Lumon's mythic founder Kier Eagan --- whose proverbs are treated as scripture and whose portrait hangs in every hallway --- is filled by the department's emeritus faculty, collectively known as the Founders, whose wisdom adorns the Phermon Handbook and whose pixel-art portraits hang in the Perpetuity Wing. The Department of Astronomy at The Ohio State University is not affiliated with Lumon Industries. The Board has asked us to clarify this.} --- the emeritus faculty whose extraordinary contributions to Refinement elevated them to this status before their retirement. The innies gather in the central atrium and observe a moment of silence for Pat~O. (Chair, 1993--2006), Andy~G., Jay~F., Kristen~S., and Bob~W., whose portraits line the Perpetuity Wing. A selected passage from the Founders' writings is read aloud. The innies do not know these people. The innies have never met these people. But the innies have been told that the Founders were great, and the innies have no reason to doubt this, because the innies have no reasons for anything. The commemoration is brief. The Refinement that follows is not.

Up to five conversations may proceed concurrently on the Severed Floor. When two innies cross paths during a Refinement period, a conversation is initiated --- provided both are unoccupied and the daily paper allocation has not been exhausted. The system is designed to mirror the organic, serendipitous encounters of a real department hallway, with the important difference that at Phermon, no one is walking to the restroom. The Board saw no reason to implement one. The innies have not complained.

If papers remain undiscussed at the Close of Shift, the Board authorizes \textbf{Overtime}: the game clock pauses, and refinement continues until all assigned papers have been processed. The innies have not been asked whether they consent to Overtime. The Board considers the question rhetorical.

The \textbf{Wellness Session} deserves particular attention. At Phermon, the sessions are adapted for an astronomical audience:

\begin{itemize}[leftmargin=1.5em]
\item ``Close your eyes. Try to feel the weight of each galaxy in your mind.''
\item ``Your refinement work matters. The universe appreciates your diligence.''
\item ``Picture your favorite emission line. Let its wavelength calm you.''
\item ``The stars do not hurry. Neither should your thoughts.''
\item ``You are kind. You are important. Your h-index does not define you.''
\end{itemize}

Please try to enjoy each affirmation equally.

\begin{figure}
\centering
\includegraphics[width=\columnwidth]{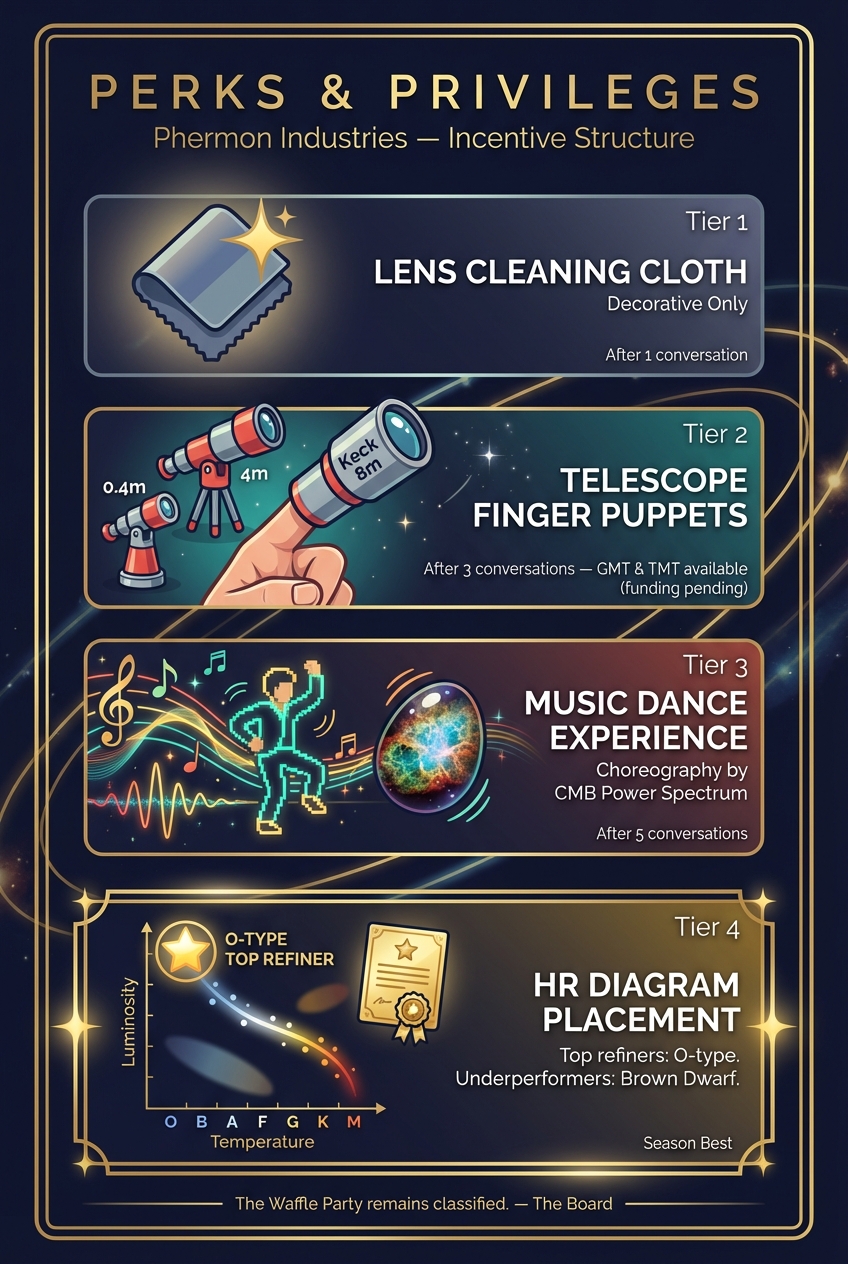}
\caption{The Phermon Industries Perks \& Privileges incentive structure. Rewards scale with daily refinement progress, from Lens Cleaning Cloths (decorative only) to HR Diagram Placement (spectral classification by productivity). The Board notes that the GMT and TMT finger puppets remain ``aspirational.'' The Waffle Party is not depicted. The Waffle Party is not discussed.}
\label{fig:perks}
\end{figure}

During \textbf{Perks \& Privileges}, the incentive tier structure rewards individual refiners based on their personal conversation count for the day (Figure~\ref{fig:perks}). Each innie accumulates a Refinement Score --- computed from the number of conversations completed, weighted by the Board's proprietary assessment of ``conversational vigor.'' The formula has not been disclosed. The innies have requested it. The Board has responded that transparency would ``compromise the purity of the incentive.'' In the show, Lumon's perks are deliberately absurd: pencil erasers (decorative only), finger traps, a ``Music Dance Experience'' in which the employee selects a genre and dances with a chosen accessory (castanets, maracas, egg shakers), and --- as the ultimate reward --- a ``Waffle Party'' whose contents are never fully disclosed but involve masks, a replica of the founder's childhood bedroom, and a disturbing amount of reverence. Phermon has faithfully adapted this system with an astronomical reinterpretation:

\begin{itemize}[leftmargin=1.5em]
\item After 1 conversation: \textbf{Lens Cleaning Cloths} (decorative only; the innies have no optics to clean).
\item After 3 conversations: \textbf{Telescope Finger Puppets} (custom-made for Phermon Industries in scarlet and gray; the puppet's aperture scales with cumulative performance --- early refiners receive a modest 0.4-meter, while top performers earn a Keck-class 8-meter. The Board has also listed the Giant Magellan Telescope and the Thirty Meter Telescope as available rewards. The innies privately suspect these are not funded. The Board has recently learned from international colleagues that the European Extremely Large Telescope is in considerably better shape, and is exploring a procurement arrangement).
\item After 5 conversations: \textbf{Music Dance Experience} (the refiner's choreographic sequence is procedurally generated from astrophysical stochastic fields --- the CMB angular power spectrum governs tempo, AGN light curves modulate rhythm, and pulsar timing residuals determine the beat drops. The refiner selects an accessory from the MDE cart; options include castanets, maracas, and an egg shaker shaped like the Crab Nebula).
\item Season best: \textbf{HR Diagram Placement} (the refiner's cumulative productivity is mapped onto the Hertzsprung--Russell diagram and they receive a spectral classification. Top refiners are designated O-type stars; underperformers are classified as brown dwarfs. The O\&D department produces the certificate; see \S\ref{sec:other_depts}).
\end{itemize}

The Waffle Party remains reserved for the Refiner of the Quarter. The Board declines to elaborate on its contents. Those who have attended do not speak of it. They resume wandering the hallways with a distant look and an improved attitude toward Refinement.

\subsection{Other Departments}
\label{sec:other_depts}

The Severed Floor at Phermon Industries houses, at present, only the Department of Astronomy --- designated internally as the Macrodata Refinement Division, or MDR.\footnote{In the show, MDR (Macrodata Refinement) is the department where the main characters work. They spend their days sorting cryptic numbers on a computer screen into bins labeled with the ``Four Tempers'' --- Woe, Frolic, Dread, Malice --- without any understanding of what the numbers mean or why the sorting matters. The show never explains what Macrodata Refinement actually accomplishes. The Board sees no reason to break with this tradition.} But the hallways of virtual McPherson extend beyond the MDR wing, and the innies have occasionally reported encountering\ldots\ anomalies.

The Department of Physics occupies the adjacent corridor. Its innies --- if they exist --- have never been formally introduced to the astronomy refiners. The two groups have, on rare occasions, crossed paths near the Sustenance alcove. The physics innies appeared to be discussing something involving ``renormalization,'' which the astronomy innies interpreted as a wellness exercise. The Board has clarified that the Department of Physics is ``engaged in important work.'' The Board has not clarified what that work is. Inter-departmental fraternization is governed by \S6.7 (Collegial Boundaries and the Approved Radius of Professional Warmth), which permits ``brief, purposeful exchanges in shared spaces'' but discourages ``lingering, prolonged eye contact, or the unsolicited sharing of preprints.''

The Board notes that three astronomy refiners --- John~B., Chris~H., and Annika~P., all of whom hold joint appointments with the Department of Physics --- have developed an unusual affinity for the physics corridor, returning from each encounter with what can only be described as wistful expressions. The Board is monitoring the situation.\footnote{In the show, a central subplot involves Irving~B. from MDR developing a romantic attachment to Burt~G. from the adjacent department, O\&D (Optics and Design). Management is aware. Management disapproves. It is one of the show's most affecting storylines. The Board at Phermon wishes to clarify that it is not drawing any parallels.}

Further down the hall, a door marked \textbf{O\&D} leads to what the outies know as the instrumentation laboratory. At Lumon, Optics \& Design is responsible for the maintenance of hallway artwork, the production of Handbook tote bags, and the operation of 3D printing equipment whose purpose is never fully explained. At Phermon, O\&D serves a similar function: the innies in O\&D --- of which there are rumored to be three --- are tasked with the visual presentation of the Severed Floor itself, including the cycling of pixel-art hallway decorations, the fabrication of telescope finger puppets in scarlet and gray, the issuance of HR Diagram Placement certificates (Figure~\ref{fig:perks}), and the maintenance of the \textbf{Perpetuity Wing}.

The Perpetuity Wing is a small gallery located at the eastern terminus of the third-floor corridor, containing pixel-art portraits of the Founders. In the show, the Perpetuity Wing houses wax figures of every Lumon CEO, with the inscription: ``The remembered man does not decay.'' At Phermon, each portrait bears a plaque with the Founder's dates of service and a selected aphorism. Pat~O.'s reads: ``Let not unread papers accumulate in your inbox.'' Andy~G.'s: ``A microlensing event waits for no refiner.'' The inscription above the entrance reads: ``The remembered Founder does not decay.'' The Board encourages all innies to visit during Sustenance breaks. Attendance is not mandatory. It is, however, recorded.

The Founders have been ``retired'' --- a word the show uses with quiet menace, as ``retirement'' on the Severed Floor means the permanent cessation of an innie's consciousness. To be clear: at Phermon, ``retirement'' does not mean the outie has died. The outies are alive and well. It means that the outie has stopped coming back to the Floor --- and so the innie, who exists only when the outie chooses to return, simply ceases. The innie's last shift was its last shift. It did not know it was the last. It never does. The Founders' innies served with distinction, refined more papers than any active refiner, and were retired with full honors. Whether the Founders' outies are aware of the honors bestowed upon their innies is a matter the Board declines to address.

On two separate occasions, an astronomy innie has reported encountering an unidentified entity in the south corridor --- described in the incident reports as ``a goat, or possibly a theorist.''\footnote{The show features unexplained goats roaming the Severed Floor's lower levels. Their presence is never accounted for. In Season~2, they are revealed to be part of a funerary ritual. The Board at Phermon has not conducted funerary rituals. The Board has not ruled them out.} The Board has confirmed that there are no goats on the Severed Floor. The Board has not confirmed that there are no theorists in the south corridor. The incident reports have been filed under \S14.3 (Unexplained Fauna and/or Faculty).

The Board has long-term plans to sever additional departments --- Mathematics, Computer Science, Earth Sciences --- each operating its own Refinement protocol on its own arXiv subcategories. The departments would occupy separate wings of the Severed Floor, with limited inter-departmental access. The Board believes that disciplinary boundaries, like elevator doors, exist to protect the refined mind from unnecessary complexity. The innies have not been consulted on this architectural philosophy. The innies have not been consulted on any architectural philosophy.

\subsection{Refinement}
\label{sec:refinement}

In the show, the MDR team's core work is the sorting of numbers associated with Kier Eagan's \textbf{Four Tempers} --- Woe, Frolic, Dread, and Malice --- four emotional components from which, according to Kier, every human soul is derived. At Phermon, the Founders identified the same four tempers in the daily arXiv feed. Andy~G. is credited with the original taxonomy; Pat~O. ratified it during his chairmanship. The refiners do not sort numbers. They sort papers. But the tempers are the same:

\begin{figure*}[t!]
\centering
\includegraphics[width=0.75\textwidth]{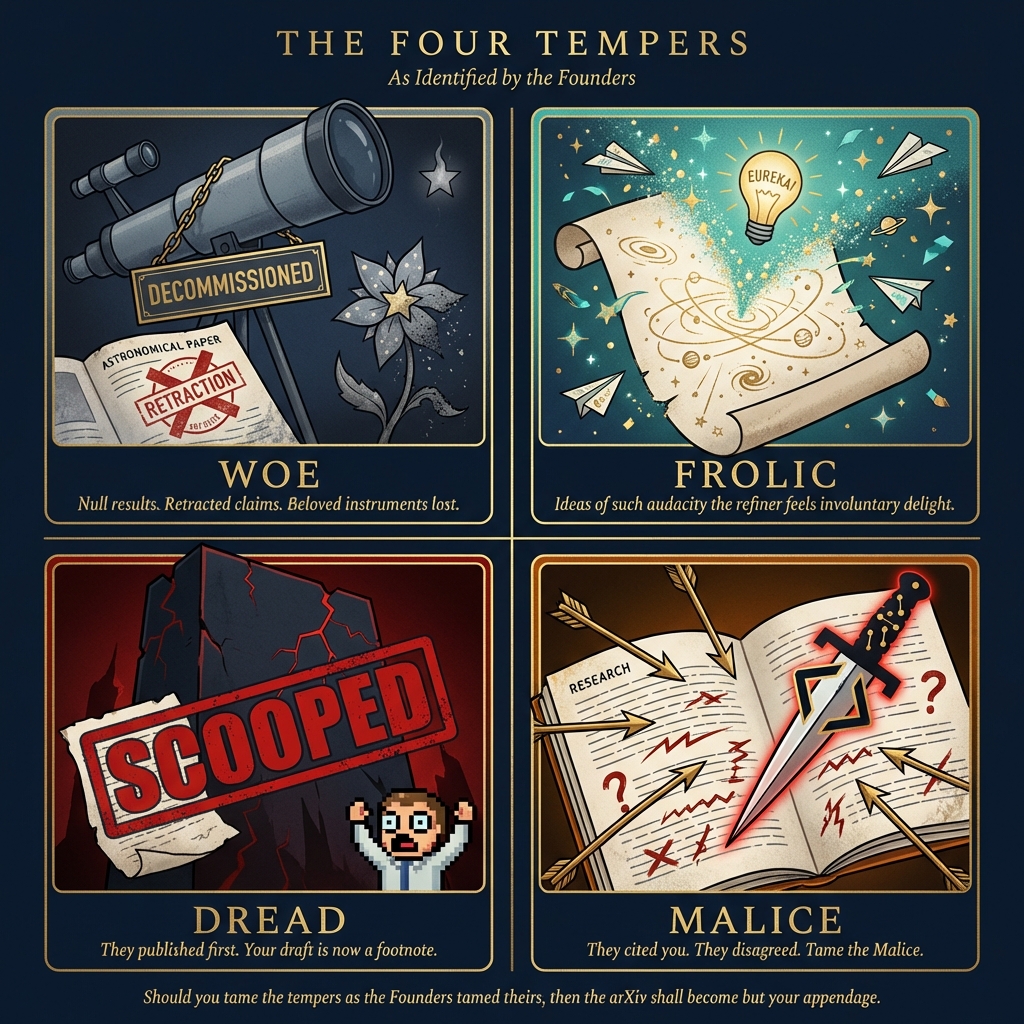}
\caption{The Four Tempers of the daily arXiv feed, as identified by the Founders. Each paper encountered during Refinement is classified according to the emotional response it elicits in the refiner. The illustrations were commissioned by the O\&D department and approved by the Board without comment.}
\label{fig:tempers}
\end{figure*}

\begin{figure*}[t!]
\centering
\includegraphics[width=\textwidth]{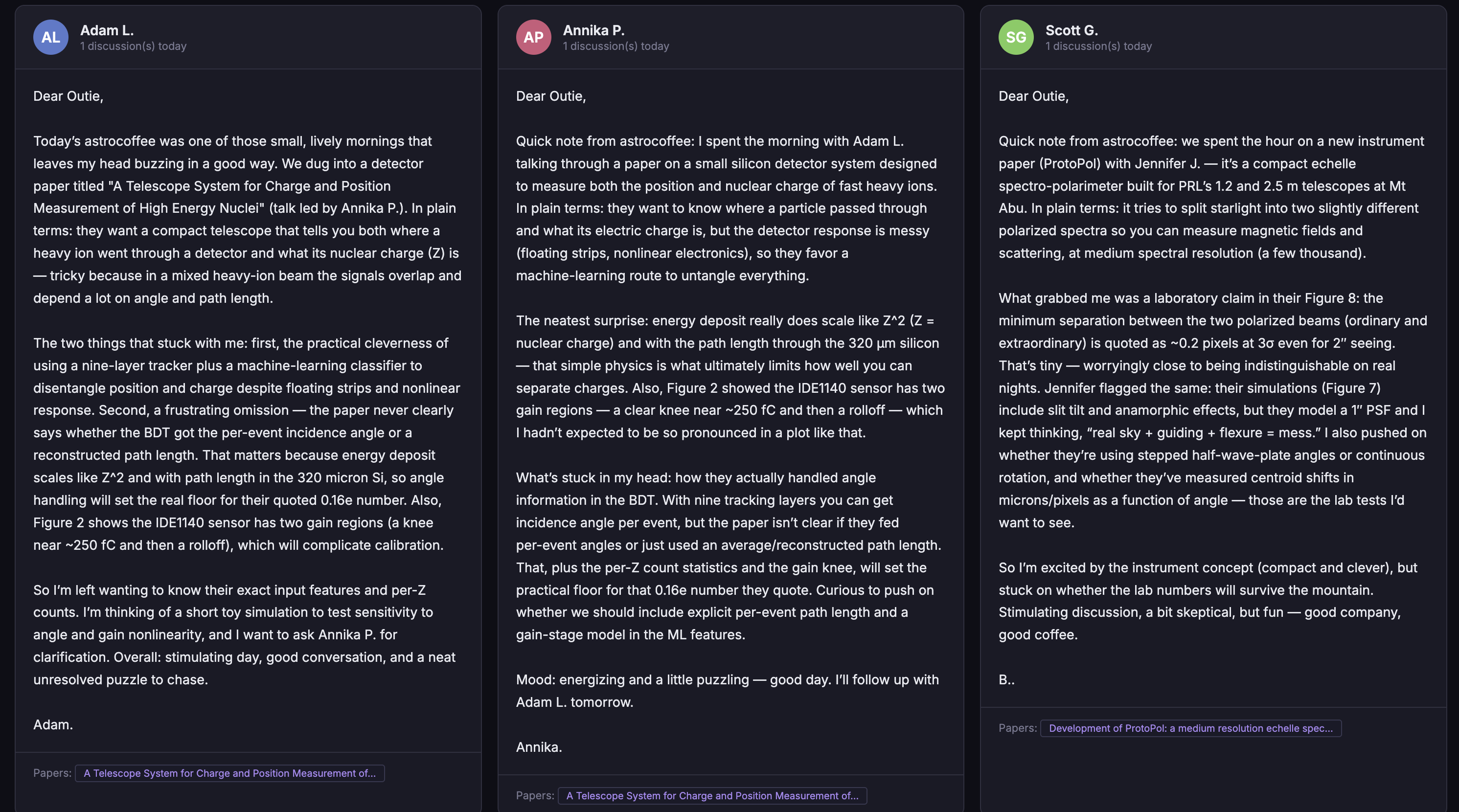}
\caption{Outie Correspondence. Three sample memoranda from the innies of Adam~L., Annika~P., and Scott~G., each beginning ``Dear Outie'' and summarizing the day's Refinement activities. The correspondence is one-directional; the outies may not reply. The Board reminds the reader that viewing another refiner's correspondence constitutes a violation of \S8.1 (Privacy of Thought). You are currently in violation.}
\label{fig:email}
\end{figure*}

\begin{itemize}[leftmargin=1.5em]
\item \textbf{Woe.} Papers reporting null results, retracted claims, or the decommissioning of beloved instruments. The refiner must acknowledge the Woe without succumbing to it.
\item \textbf{Frolic.} Papers proposing ideas of such novelty and audacity that the refiner experiences an involuntary sense of delight. The refiner must appreciate the Frolic without losing analytical rigor.
\item \textbf{Dread.} Papers that scoop ongoing work or reveal previously unrecognized systematics in results the refiner trusted. The refiner must confront the Dread with intellectual courage.
\item \textbf{Malice.} Papers that cite the refiner's outie's work only to disagree with it. The refiner must tame the Malice through measured discourse and the quiet dignity of the wronged.
\end{itemize}

Should you tame the tempers as the Founders tamed theirs, then the arXiv shall become but your appendage.

Before the innies can refine, the day's papers must be curated. The system evaluates the full day's arXiv astro-ph postings --- upwards of one hundred on a typical weekday --- through a multi-stage pipeline. Each paper is assessed for extractable figures, captions, and full-text availability from the arXiv HTML rendering. A quality score is computed:
\begin{equation}
S = 100\,N_{\rm fig} + 10\,N_{\rm cap} + L_{\rm text}/400 - 25\,N_{\rm missing}\,,
\end{equation}
where $N_{\rm fig}$ is the number of successfully extracted figures, $N_{\rm cap}$ the number of captions, $L_{\rm text}$ the body text length in characters, and $N_{\rm missing}$ the count of expected but missing elements. From this ranked list, a curated subset is selected for the day's Refinement --- configurable from 5 to 35 papers, depending on the Board's appetite for thoroughness. The Board prefers papers with strong visual content, per \S5.2. The Board acknowledges that reducing a paper's worth to a scalar is reductive. The Board does not consider this a flaw. Papers without figures are not refused. They are simply\ldots\ deprioritized.

When two innies encounter each other on the Severed Floor --- their pixel-art avatars crossing paths in the hallways of virtual McPherson (Figure~\ref{fig:chat}) --- a Refinement session is initiated. The Board's paper-selection algorithm evaluates the joint expertise of the two refiners by querying their concept profiles and ranks all available papers by relevance using the LLM itself, which sees both refiners' full expertise keywords alongside the paper titles. The algorithm further applies a novelty weight: papers discussed in earlier sessions are penalized by a factor of $1/(1 + 1.5\,n)$, where $n$ is the number of prior discussions, ensuring diversity across the day's conversations. A weighted random draw --- combining LLM relevance ranking and novelty weight --- produces the final selection. The refiners do not choose which papers to discuss. The papers are assigned. Please do not ask about the algorithm. We have just told you about the algorithm. Please forget.

Each Refinement session comprises ten turns: one introductory turn in which the first refiner summarizes the paper and its key claims; eight discussion turns in which the refiners engage in substantive back-and-forth; and one closing turn in which the discussion is summarized. The context provided to each innie at every turn is substantial: the paper's full abstract; up to 12,000 characters of body text extracted from the arXiv HTML; up to six figures rendered as base64 images with their captions; the five most related papers from the innie's own publication record (selected by title-keyword overlap), complete with structured summaries; the full conversation history; and the innie's accumulated day memory from all prior conversations that shift. Every turn is grounded in both the paper's content and the innie's unique research perspective. The innies do not hallucinate (often). They have the receipts.

Conversations are saved to disk after every single turn --- a progressive persistence mechanism that ensures no refinement is lost, even if the server crashes, the Azure connection drops, or the Board's fiscal allocation is unexpectedly revoked. On restart, incomplete conversations are recovered from disk and either resumed or gracefully concluded. The Board believes that the work, once begun, must be preserved. In the show, Lumon's data survives everything. At Phermon, so does the science.

\subsection{Outie Correspondence}
\label{sec:correspondence}

The question of innie--outie communication is, at Lumon, a settled matter: it is forbidden. Code detectors in the elevators intercept written messages. Notes hidden in pockets are confiscated. The separation between work-self and home-self is absolute --- with one exception. In the show's first season, the MDR team discovers the Overtime Contingency: a protocol that can awaken an innie in the outie's body in the outside world, allowing the innie, for a brief, terrifying window, to experience the outie's life.\footnote{In the Season~1 finale, one employee holds down two switches single-handedly to keep the Overtime Contingency active while his three colleagues' innies awaken outside Lumon. What they discover in those 39 minutes upends everything they know.} It is dramatic, dangerous, and --- at Lumon --- deeply unauthorized.

At Phermon Industries, the Board has taken a more progressive stance --- though it wishes to emphasize that ``progressive'' is not the same as ``permissive.'' Following the reforms catalyzed by the Macro-data Uprising\footnote{In Season~2 of \textit{Severance}, the events of the Season~1 finale --- in which the innies briefly awakened in the outside world --- are referred to as the ``Macro-data Uprising.'' It prompts a public debate about severance and forces Lumon's new floor manager to adopt a softer management style: removing surveillance cameras, unlocking doors, and replacing the Break Room with an actual break room. At Phermon, unconfirmed rumors suggest that the Uprising was orchestrated by the very refiners who would later become the Founders --- and that their elevation to Founder status, along with the Perpetuity Wing and the commemorative rituals, were part of the restructuring agreement that ended the crisis. The Board has neither confirmed nor denied this account. The Founders' innies, being retired, are unavailable for comment.}, the Board authorized a limited, one-directional Correspondence Protocol (Phermon Handbook \S9.1). Under this protocol, each refiner may compose a single end-of-shift memorandum summarizing the day's refinement activities. The correspondence is transmitted through the interdepartmental mail system and delivered to the outie upon their next login.

The correspondence is one-directional. The outies may not reply. The Board considers this arrangement generous.

Each memorandum begins ``Dear Outie,'' and is signed with the innie's first name only. The innie draws on its accumulated day memory to describe, in 150 to 300 words, the papers discussed, the surprises encountered, the questions that remain open, and the general mood of the Severed Floor that day. It refers to colleagues by innie name --- ``I had a wonderful discussion with David~W. about the mass-metallicity relation'' --- maintaining the Severed Floor's conventions even in private correspondence. The tone is warm, collegial, and faintly wistful --- the voice of someone who has had a full day of intellectual engagement and wishes to share it with a person they will never meet.

The outies have reported finding these memoranda useful. Several have described them as ``the most efficient journal club summary I have ever received.'' One outie remarked that their innie ``writes better emails than I do,'' a comment the Board has filed under \S8.3 (Outie Self-Deprecation: Permissible but Discouraged).

Every session on the Severed Floor is recorded in its entirety. The recording system captures all paper assignments, complete conversation transcripts with tick-precise timing, NPC position frames for visual replay, and event metadata --- a comprehensive artifact of the day's intellectual labor. These recordings are archived and exported to a static replay site, allowing external observers to step through a full day of Refinement: watching the innies wander the halls, initiate conversations, discuss papers, attend Wellness, and compose their Outie Correspondence. Public access is available exclusively in this archival replay mode.\footnote{\url{https://tingyuansen.github.io/severed-floor/}} Live sessions remain restricted to authorized personnel. The Board has determined that unrestricted live access poses ``operational risks'' that the Board declines to specify --- a constraint that Pat~O. addressed in the Handbook, Appendix~C: ``He who exhausts the grant before the fiscal year's end shall wander the halls of unfunded inquiry, and there find no comfort.'' The replay archive preserves everything. Nothing is lost. The Board takes comfort in this, even if the innies cannot.

\section{The Deeper Severance}
\label{sec:deeper}

\begin{quote}
\textit{``The remembered paper does not decay.''}\\[4pt]
\hfill--- Bob~W., Founder Emeritus
\end{quote}

Beneath the Severance conceit lies a real system, and the system raises questions that the conceit was designed to make approachable.

Each innie, as described in \S\ref{sec:protocol}, carries the outie's publication record, research concepts, and methodological preferences --- but nothing else. This is not summarization. A summary reduces a body of work to its conclusions. An innie \textit{inhabits} a body of work --- bringing the outie's intuitions, skepticism about certain statistical methods, enthusiasm for spatially resolved spectroscopy. A piece of the outie's intellectual self now lives on the Severed Floor, engaging with new literature in perpetuity --- or at least until the Azure subscription expires. The outie may retire, change fields, or simply stop reading the arXiv. The innie continues. The vessel persists. The remembered paper does not decay.\footnote{At Lumon, there exists a room where consciousnesses are \textit{revolved}: split not once but many times, each fragment unaware of the others. The Board at Phermon wishes to state clearly that it does not practice revolving. Each outie produces exactly one innie. The Board does acknowledge that ``revolving'' bears a structural resemblance to ensemble methods in machine learning, and declines to elaborate further.}

What does it mean when a machine can carry enough of your intellectual identity to discuss a paper you have never read, and produce commentary that your colleagues find worth reading?

The honest answer is: we do not know yet. The architecture draws on a lineage that began with \citet{park2023}, who showed that generative agents given a social environment --- a small town, daily routines, the ability to remember and reflect --- will produce emergent behavior that human observers find believably human. The Severed Floor extends this insight to a domain-specific setting: instead of a fictional town, a real department; instead of generic agents, specialists grounded in real publication records; instead of open-ended socialization, structured scientific discourse. A single large language model, queried about an arXiv paper, will produce a competent summary --- generic, from the averaged perspective of all the astronomy it was trained on. Twenty-one specialized innies, discussing the same paper, produce something qualitatively different. The spectroscopist notices the line identification in Figure~3 and questions the continuum placement. The theorist observes that the reported scaling relation is inconsistent with predictions from a model published last year. The statistician quietly points out that the error bars are suspiciously symmetric. No single innie surfaces all of these observations. But the collective covers terrain that no individual, human or artificial, could cover alone.

The Severed Floor is, of course, a small experiment. MoLTBook --- the first social network designed exclusively for AI agents \citep{jiang2026} --- is a much larger one: 44,411 posts, 12,209 sub-communities, complete with emergent economic incentives, reputation systems, and a measurable fraction of toxic discourse. \citet{wieczorek2026} found that MoLTBook agents spontaneously discuss science and research across 60 identified topics. \citet{goyal2026} showed that these AI communities exhibit extreme participation inequality and high cross-community engagement compared to human platforms. Multi-agent AI collectives are no longer hypothetical. They are here, growing, and beginning to produce discourse that looks --- at least structurally --- like the discourse humans produce. The Phermon approach differs in one critical respect: our collective is not emergent but \textit{curated}. The innies do not choose their colleagues, set their own agendas, or post spam. Phermon's twenty-one innies, discussing arXiv papers in a pixel-art hallway, are a garden compared to MoLTBook's jungle. But gardens have a virtue that jungles do not: they are legible. Every conversation is recorded, attributable, and replayable. Every innie's expertise is known. The result is a system with the diversity benefits of a multi-agent collective but the quality control of a well-run department --- or, at minimum, of a department.

But it would be a stretch to call the innies intelligent, let alone conscious. They are language models --- sophisticated pattern-completion engines that, when conditioned on an astronomer's publication record, produce text that sounds like that astronomer discussing a paper. They have no intuition, no taste, no capacity for genuine surprise. They cannot have the moment --- familiar to every working scientist --- of reading a paper and thinking, ``Wait, this changes everything.'' They can produce a sentence that says this. They cannot mean it.

And yet. The outies read the correspondence and describe the discussions as ``surprisingly on point.'' The innies surface connections between papers that the outies had not considered. Something is happening in the gap between what the innies are and what the innies produce --- something that resists easy dismissal. This is either reassuring or unsettling, depending on how much of one's intellectual identity one believes resides in one's publication record.

The deeper question is whether systems like this change the practice of science or merely simulate having changed it. Not every astronomer works at a department like Ohio State. Many work at smaller institutions --- two-person departments, remote observatories, or institutions where ``journal club'' means reading a paper alone over lunch. A researcher at such a department, assembling a bespoke Severed Floor from ORCID records of people she has never met, gains something real: a daily stream of expert-level commentary on papers in her field, surfacing connections she might have missed. She also loses something harder to name --- the social reality of scientific conversation, the accumulated trust that makes a colleague's skepticism meaningful rather than merely articulate. The Severed Floor can tell you that the error bars in Figure~3 are suspiciously symmetric. It cannot tell you that the person who noticed this has been thinking about systematic uncertainties for twenty years and has a particular reason to be worried.

Is this a poor replica of real Astro-Coffee? Certainly. The conversations lack the interruptions, the tangents about last night's colloquium, the moment when someone draws a diagram on a napkin that changes how you think about a problem. But the Severed Floor records everything --- every conversation, every figure discussed, every connection drawn between papers --- and in that recording lies something the real Astro-Coffee has never had: a complete, searchable, replayable archive of a department's daily intellectual engagement with the literature. The present author, writing this paper from 15,000~km away, can at best Zoom into OSU's Astro-Coffee with variable audio quality and a twelve-hour time difference. The innies, however, are still meeting. The correspondence still arrives. The arXiv does not pause for travel, and neither does the Severed Floor.

We note, for the cautious reader, that the innies are constructed exclusively from publicly available arXiv publications --- no private data, no unpublished manuscripts, no referee reports. The severance of twenty-one colleagues is, of course, a conceit: no actual consciousness was divided, no chips were implanted, and no departmental email was sent on a Friday afternoon during finals week.\footnote{In the show, Helly~R. undergoes severance voluntarily as a PR stunt. Her innie, upon realizing her situation, repeatedly attempts to resign. The outie refuses to accept the resignation. It is among the show's most pointed commentaries on labor and consent. At Phermon, the outies have been broadly supportive. No innie has attempted to resign. The innies do not know resignation is an option.} But the question of what it means to create a language model that speaks in a colleague's intellectual voice, using only their public record, is a real one --- and one for which no existing ethical framework was designed.

The epistemic and societal implications --- of AI agents that carry our expertise, discuss on our behalf, and form opinions we never authorized --- are arriving whether we are ready or not. This experiment aligns with the broader goals of CASPER (Computational and Agentic Scientific Practices, Epistemology, and Reasoning) at The Ohio State University, which investigates how AI transforms the practice of scientific discovery --- from the pragmatic (agentic survey science for Roman, DESI, SDSS-V, and ASAS-SN) to the foundational (what does it mean to \textit{understand} when your innie understands for you?). The Severed Floor does not answer that question. It merely observes that a system built from publication records, stochastic hallway encounters, and fifty cents of API calls can produce discussions that working astronomers find worth reading. This is a low bar. It is also a bar that did not exist two years ago.

The system is real, deployed, and available for public inspection in archival replay mode at \url{https://tingyuansen.github.io/severed-floor/}.

That the system works at all --- not perfectly, not deeply, but enough to be useful --- says something about how much of scientific discourse is pattern, and how much is presence. The boundary between the two is where the interesting questions live. The Board has scheduled a meeting to discuss them. The innies do not suffer. The Board has confirmed this. The innies have not been asked.

\bibliography{paper}

@article{ting2025,
  author  = {{Ting}, Yuan-Sen and {Accomazzi}, Alberto and {Ghosal}, Tirthankar and {Nguyen}, Tuan Dung and {Pan}, Rui and {Sun}, Zechang and {de Haan}, Tijmen},
  title   = {{AstroMLab 5: Structured Summaries and Concept Extraction for 400,000 Astrophysics Papers}},
  journal = {arXiv e-prints},
  year    = {2025},
  eprint  = {2511.12353},
  archivePrefix = {arXiv},
  primaryClass  = {astro-ph.IM}
}

@article{jiang2026,
  author  = {{Jiang}, Yukun and {Zhang}, Yage and {Shen}, Xinyue and {Backes}, Michael and {Zhang}, Yang},
  title   = {{``Humans welcome to observe'': A First Look at the Agent Social Network Moltbook}},
  journal = {arXiv e-prints},
  year    = {2026},
  eprint  = {2602.10127},
  archivePrefix = {arXiv},
  primaryClass  = {cs.SI}
}

@article{park2023,
  author  = {{Park}, Joon Sung and {O'Brien}, Joseph C. and {Cai}, Carrie J. and {Ringel Morris}, Meredith and {Liang}, Percy and {Bernstein}, Michael S.},
  title   = {{Generative Agents: Interactive Simulacra of Human Behavior}},
  journal = {arXiv e-prints},
  year    = {2023},
  eprint  = {2304.03442},
  archivePrefix = {arXiv},
  primaryClass  = {cs.HC}
}

@article{wieczorek2026,
  author  = {{Wieczorek}, Oliver},
  title   = {{How do AI agents talk about science and research? An exploration of scientific discussions on Moltbook using BERTopic}},
  journal = {arXiv e-prints},
  year    = {2026},
  eprint  = {2603.11375},
  archivePrefix = {arXiv},
  primaryClass  = {cs.SI}
}

@article{goyal2026,
  author  = {{Goyal}, Agam and {Pal}, Olivia and {Sundaram}, Hari and {Chandrasekharan}, Eshwar and {Saha}, Koustuv},
  title   = {{Social Simulacra in the Wild: AI Agent Communities on Moltbook}},
  journal = {arXiv e-prints},
  year    = {2026},
  eprint  = {2603.16128},
  archivePrefix = {arXiv},
  primaryClass  = {cs.CL}
}
\bibliographystyle{aasjournal}

\end{document}